\documentclass[final,english]{bullsrsl}[2022/06/15]

\usepackage[latin1]{inputenc}
\usepackage[T1]{fontenc}

\usepackage{natbib} 
\usepackage{graphicx}
\usepackage{amssymb}

\begin{document}
\title{Multiwavelength view of massive binaries}

\author[affil={1},corresponding]{Bharti}{Arora}
\author[affil={1}]{Micha\"{e}l}{De Becker}
\author[affil={2}]{Jeewan C.}{Pandey}
\affiliation[1]{Space Sciences, Technologies and Astrophysics Research (STAR) Institute, University of Li\`ege, Quartier Agora, 19c, All\'ee du 6 A\^out, B5c, B-4000 Sart Tilman, Belgium}
\bigskip
\affiliation[2]{Aryabhatta Research Institute of Observational Sciences, Nainital-263 002, India}
\correspondance{bhartiarora612@gmail.com}
\date{21st May 2023}
\maketitle


%

\begin{abstract}
The high luminosity of massive, early-type stars drives strong stellar winds through line scattering of the star's continuum radiation. Their momenta contribute substantially to the dynamics and energetics of the ambient interstellar medium in galaxies. 
The detailed multi-wavelength study of massive O-type and Wolf-Rayet binaries is essential to explore the hydrodynamics of the shocks formed in the stellar outflows and wind structure. Further, deep analysis of some of the interesting phenomena like particle acceleration and dust formation associated with hot stars' winds provides a global view of stellar outflows. In this context, a few massive binaries have been explored using photometric and spectroscopic measurements in different wavebands. This paper highlights important insights gained from investigating massive binaries with several ground and space-based facilities.
\end{abstract}

\keywords{Early-type stars, Massive binaries, Colliding-winds, Multiwavelength emission}

\section{Introduction}
The stars with an initial mass several times greater than that of the Sun, typically $\geq$8 M$_\odot$,  are referred to as massive stars. A crucial feature of massive stars is their stellar winds. These are the outflow of the charged particles ejected from the surface of a star. In the case of massive stars, their winds are exceptionally strong compared to their lower-mass counterparts. Their extreme luminosity and high temperatures power these winds through line driven acceleration mechanism where an interaction between the intense ultraviolet (UV) radiation emitted by massive stars and the numerous spectral lines present in their atmospheres occurs. These lines can absorb and scatter photons, resulting in a net outward force on the surrounding gas. This mechanism is particularly important in O-type and B-type stars, which have strong UV radiation and prominent spectral lines. The powerful stellar wind injects mass, energy, and momentum into the interstellar medium (ISM). The evolution of these objects happens over short lifetimes (typically of the order of a few to 10 Myr) and appear as Wolf$-$Rayet (WR) stars during the later stages of their evolution. The massive stellar winds can have velocities of 1000$-$3000 km s$^{-1}$ and can remove a significant fraction of the star's outer layers during its lifetime with mass loss rates in the range of 10$^{-7}$ to 10$^{-4}$ M$_\odot$ yr$^{-1}$ \citep{2008A&ARv..16..209P}.

Being the evolutionary descendants of massive O stars, the WR stars expose their H or He-burning cores as a result of substantial mass loss. Spectroscopically, these stars are spectacular in appearance. Instead of the narrow absorption lines which are typical of `normal' stars, their optical and UV spectra are dominated by strong and broad emission lines. These emission lines are formed far out in the wind as both line- and continuum-emitting  regions  are  much  larger  than  the  conventional stellar radius. WR stars come in two flavours: those with strong emission lines of He and N (WN subtypes), and those  with strong  He,  C,  and O  lines  (WC and  WO subtypes). The products of the CNO cycle and triple-$\alpha$ nuclear reactions are revealed on the surfaces of WN and WC$/$WO subtypes, respectively \citep{2007ARA&A..45..177C}. 

Despite the short lifetimes of massive stars, these stars impact the ISM and the ecology of their host galaxies tremendously. The emission of huge amounts of UV photons is the dominating reason for the ionization of the ISM \citep{2004AdSpR..34...27R}. The powerful outflows from stellar surfaces substantially influence the evolution of these stars by modifying their evolutionary time scales, chemical profiles, surface abundances, and stellar luminosities. Furthermore, a huge amount of mechanical energy is transported into the ISM by these stellar winds and, also enhances the chemical enrichment of the interstellar gas significantly, during the WR stage and eventual explosive deaths as supernovae \citep{2004cmpe.conf...31L}. Therefore, it is important to study the physics of stellar winds to understand massive stars along with the quantification of their role in several interstellar feedback processes.

\begin{figure}[h]
\centering
\includegraphics[scale=0.40]{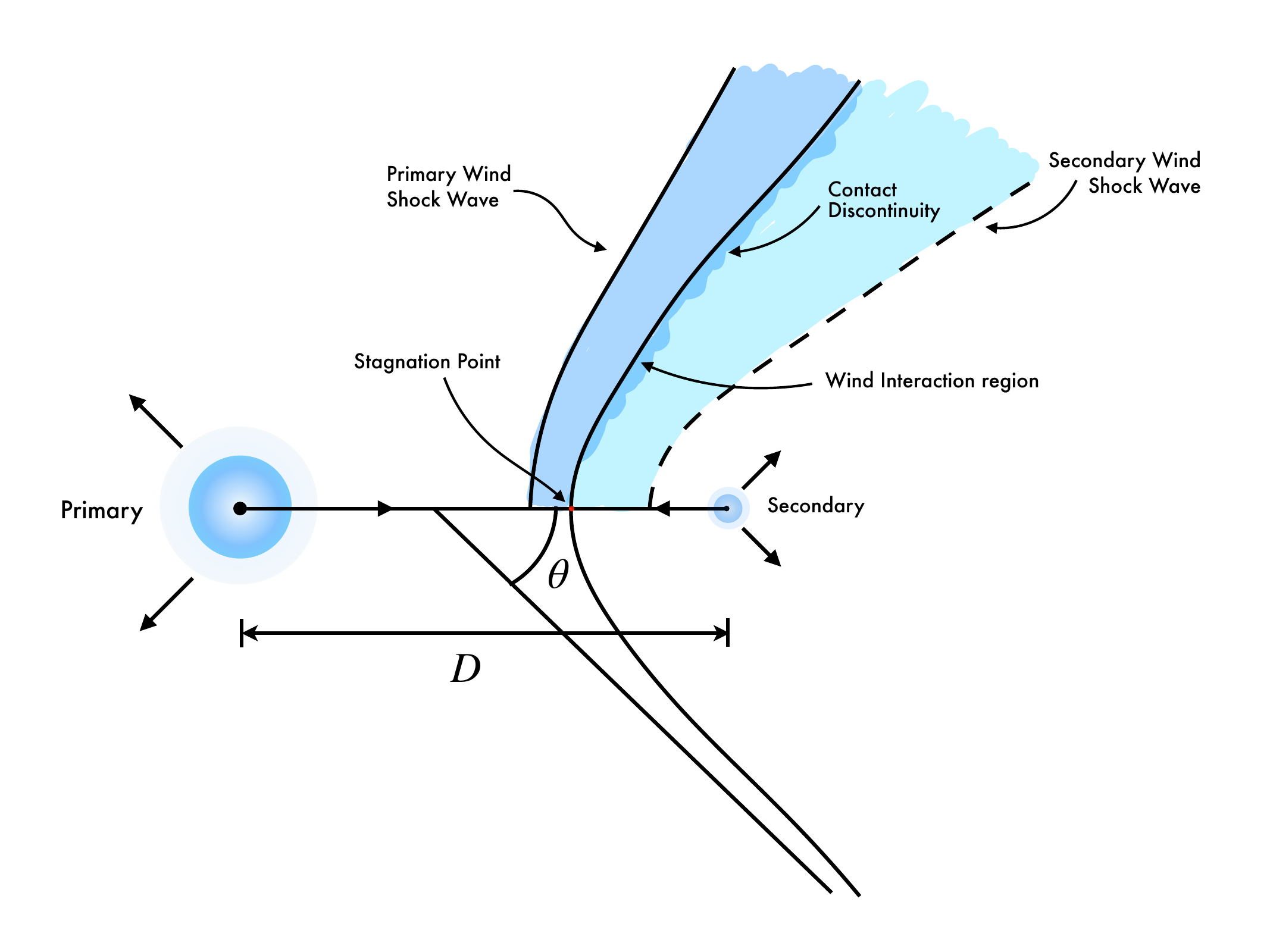}\label{fig1}

\begin{minipage}{12cm}
\caption{Schematic view of a colliding wind binary system.}
\end{minipage}
\end{figure}

Massive stars are generally found in binary or higher multiplicity systems. Conservatively, the lower limit of the fraction of massive stars in binaries is 50\% \citep{2011IAUS..272..474S}. The winds of two stars in a massive binary interact with each other and this leads to the formation of hydrodynamic shocks in the wind interaction region (WIR). The collision of stellar winds will manifest itself in the form of two oppositely oriented shock fronts. A contact discontinuity separates the two shocks in between the stars with a binary separation of `\textit{D}' as shown in Figure\,\ref{fig1}. At each shock front, part of the kinetic energy is thermalized. The location of the wind collision is determined by comparing the wind momentum of both stars. For the winds having comparable strength, WIR is located in between the two stars at an equal distance from both and is perpendicular to the line joining the center of the binary components. If the two outflows are of different strengths, the shocked gas in the collision zone is wrapped around the binary component which has relatively weaker stellar wind with a half opening angle `\textit{$\theta$}'.

This interaction is the source of many observational signatures which span over a wide range of the electromagnetic spectrum, from the radio waves to the $\gamma$-ray domain. By combining data from multiple wavelengths, a comprehensive picture of the physical processes taking place in massive binary systems can be constructed. Therefore, the multiwavelength view provides a unique way for a deeper understanding of the stellar winds, mass loss, and the associated phenomena, shedding light on the evolutionary pathways and ultimate fate of these intriguing binary systems.

\section{X-ray investigation of massive binaries}
Early type stars were first detected in X-rays by \textit{Einstein} observations of the Cyg OB2 and the Carina Nebula region (\citealt{1979ApJ...234L..51H}, \citealt{1979ApJ...234L..55S}). The shocks generated within the radiatively-driven winds of a single, early-type star are thought to emit mostly in the soft X-rays \citep{1997A&A...322..167B,1980ApJ...241..300L,1982ApJ...255..286L}. The line-driven winds are inherently not stable to Doppler perturbations. The most suitable reason is little enhancement in the speed of a packet in the outflow will push the packet out of the underlying wind material shadow. Consequently, the wind packet gets accelerated upon receiving more of the photospheric flux. Therefore, winds from a single massive star have structures and are prone to instabilities (\citealt{1988ApJ...335..914O,1997A&A...322..878F}). As a result, the massive stellar winds become clumpy and produces shocks within the wind of a single star when several clumps of different velocity collide with each other throughout the outflow. The soft X-ray emission is generated in single hot stars by these distributed shocks. The typical temperature associated with the X-ray emission from single massive stars is $\sim$ 10$^{6}$ K (suggesting pre-shock velocities of hundreds of km/s). Also, they are not significantly dependent upon time. A part of this emission originates from a distance of around two stellar radii above the photosphere or maybe from further interior parts of the wind as suggested by high-resolution X-ray spectroscopy of massive stars with high-resolution spectrographs onboard  \textit{Chandra} and \textit{XMM-Newton} (\citealt{2001A&A...365L.312K,2001ApJ...554L..55C}).

An additional source producing X-ray emission in the case of massive binary systems is the wind collision between the binary components on top of the intrinsic X-ray emission given by each star individually (\citealt{1976SvA....20....2P,1976SvAL....2..138C}). However, wind collision is strong only if the stellar winds have attained sufficiently high velocity before colliding and they are close enough to each other as well \citep{1992ApJ...386..265S}. The X-ray emission from the colliding winds can be distinguished from that of the background emission arising from the individual stars as the shocks in the wind collision are expected to have temperatures about a factor of 10 or so higher than those of the ``distributed shocks" in the individual winds. Again, the individual winds have to reach their maximum velocities  before interacting with one another. Further, the emission from the wind-wind collision zone shows high variability due to intrinsic alterations of the emitting region or modifications in the material characteristics present between the observer and the WIR or both may also affect simultaneously. The massive binaries are generally X-ray brighter when compared to the  apparently single stars \citep{1987ApJ...320..283P}.    

The long-term behaviour of a colliding wind binary (CWB) named WR 25 is explored using archival X-ray data obtained over a time span of $\sim$16 years. It is a bright (V = 8.1) WR star located in the Carina Nebula region and is classified as O2.5If*$/$WN6+OB \citep{2011MNRAS.416.1311C}. \citet{2006A&A...460..777G} studied the radial velocity profile of WR 25 and suggested that it has an eccentric binary orbit (e = 0.5). The ratio of the X-ray to the bolometric luminosity of $\sim$10$^{-6}$ \citep{1982ApJ...256..530S} for WR 25 is an order of magnitude higher than observed for single WR stars, suggesting it to be a very likely candidate of CWB systems \citep{2003A&A...402..653R,2006A&A...445.1093P}. In order to investigate this system and the associated winds, we have carried out its X-ray study using the observations made by \textit{NuSTAR}, \textit{Suzaku}, \textit{Swift}, and \textit{XMM-Newton} at 226 epochs during 2000$-$2016. This study has improved the phase coverage of the orbit of WR 25 significantly as compared to previous studies and hence enables a better understanding of the wind properties.

The thermal X-ray emission from the individual stars as well as WIR dominates below 10 keV in the massive binaries (described above). The background subtracted X-ray light curve as observed by X-ray telescope (XRT) onboard \textit{Swift} in 0.3$-$10.0 keV energy band is shown in Figure\,\ref{fig3}. Variability is clearly seen in the light curve. Long-term monitoring of WR 25 enabled us to determine its orbital period accurately. Lomb-Scargle periodogram \citep{1976Ap&SS..39..447L,1982ApJ...263..835S} was used to perform the period analysis from the \textit{Swift} light curve. The peak with the highest power is located at the frequency 0.00481\,$\pm$\,0.00005 cycles day$^{-1}$ which corresponds to 208.3\,$\pm$\,2.2 days orbital period. The estimated period is consistent with the orbital period derived by \citet{2006A&A...460..777G} which is 207.85 $\pm$ 0.02.

\begin{figure}[h]
\centering
\includegraphics[scale=0.44]{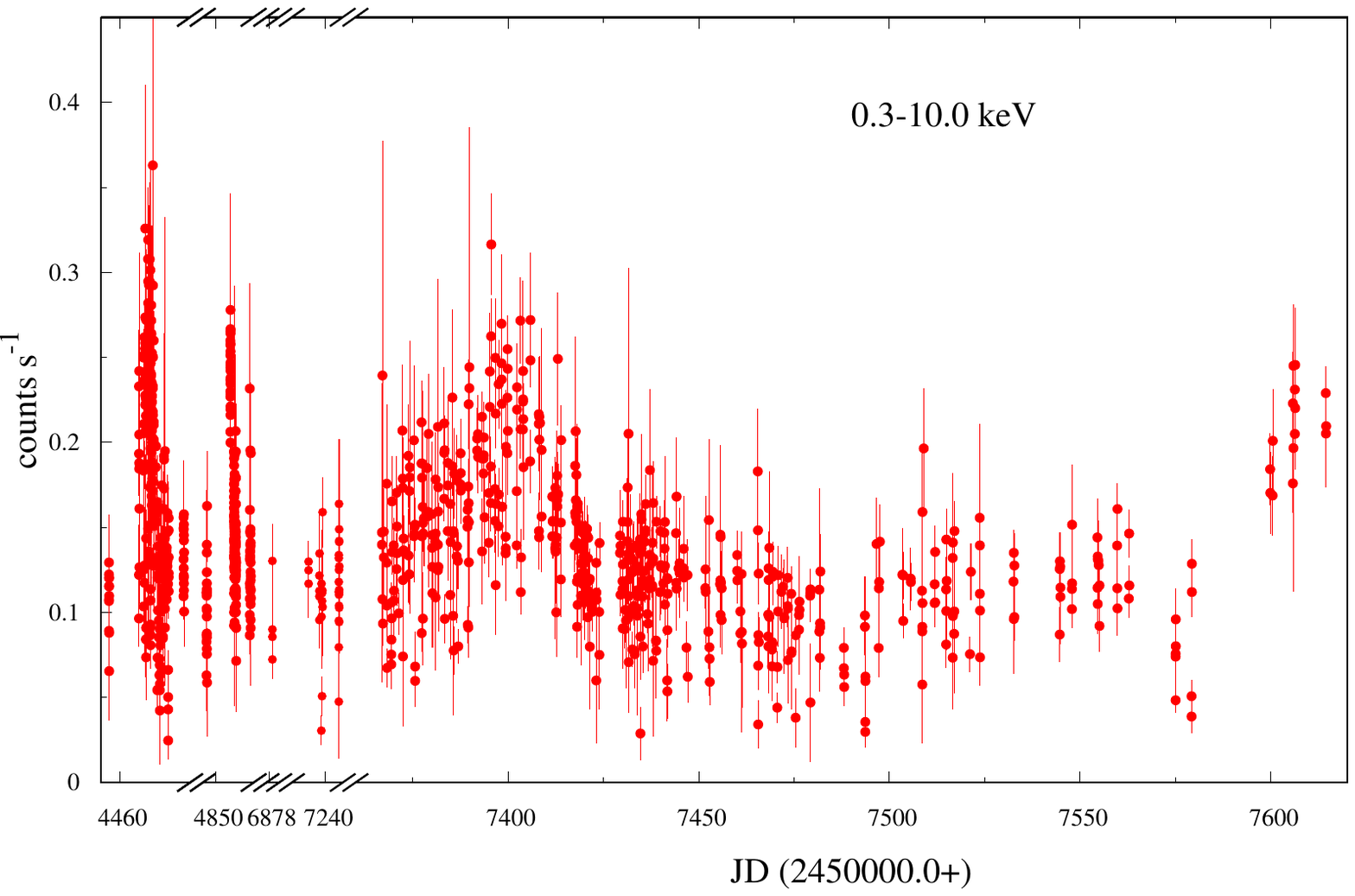}\label{fig2}
\begin{minipage}{12cm}
\caption{X-ray light curve of WR 25 as observed by \textit{Swift}--XRT.}
\end{minipage}
\end{figure}

X-ray spectra of WR 25 presented typical features of plasma heated to the temperature of 10$^6$--10$^7$ K with several emission lines. The spectra were fitted using the models of Astrophysical Plasma Emission Code (\textsc{apec}; \citealt{2001ApJ...556L..91S}) modified by the Galactic as well as local absorption effects and various spectral parameters were estimated. The variation of local hydrogen column density (N$_{H}^{local}$), ISM-corrected X-ray fluxes obtained in soft (0.3--2.0 keV, F$_{S}^{ism}$) and hard (2.0--10.0 keV, F$_{H}^{ism}$) energy bands with orbital phase is shown in Figure 3. The zero phase corresponds to the time of periastron passage of this eccentric binary. Below 10 keV, colliding stellar winds of the binary components of WR 25 result in enhanced X-ray emission as the two binary components move close to the periastron passage in both the soft and hard energy bands. This is because the wind interaction is maximum at the periastron as wind density is largest in that part of the orbit. However, it gradually becomes fainter as the two components move apart from each other close to apastron. Additionally, the enhancement in N$_{H}^{local}$ around periastron, when the line of sight passes through the denser wind of the WR star in front, creates a pronounced effect on the soft X-ray flux which is more prone to absorption than the hard X-ray photons (for details, see \citealt{2019MNRAS.487.2624A}).

\begin{figure}[h]
\centering
\includegraphics[scale=0.44]{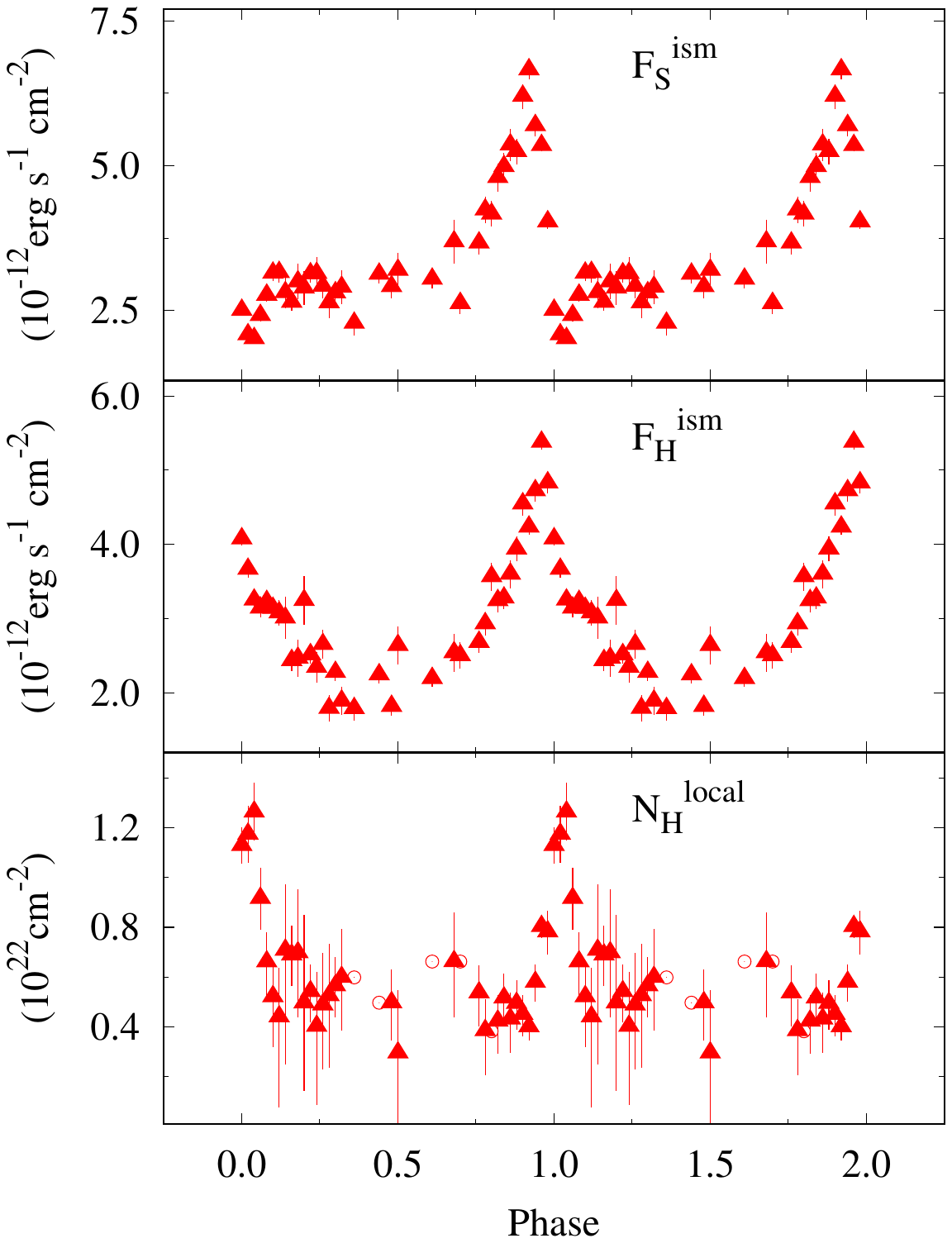}\label{fig3}
\begin{minipage}{12cm}
\caption{Variation of X-ray flux in soft (F$_{S}^{ism}$) and hard (F$_{H}^{ism}$) energy bands along with local hydrogen column density (N$_{H}^{local}$) estimated from spectral fitting of Swift-$XRT$ spectra of WR 25 in 0.3--10.0 keV energy band. The zero phase corresponds to the time of periastron passage.}
\end{minipage}
\end{figure}

Observationally, it is seen that some CWBs also act as sources of particle acceleration in their WIR through diffusive shock acceleration (DSA) mechanism which leads to the production of relativistic particles \citep{2013A&A...558A..28D}. The relativistic electrons (the major constituent of wind plasma) can inverse comptonize the photospheric stellar light to X-rays or even soft $\gamma$-rays. This opens up the possibility that some non-thermal X-ray emissions may be measured in CWBs. However, no significant X-ray emission above 10 keV was observed for WR 25 by \textit{NuSTAR} which provides evidence that no inverse Compton (IC) scattering emission is produced by WR 25 above the background level. The upper limit, derived on the putative non-thermal X-ray luminosity, is of the order of 10$^{32}$\,erg\,s$^{-1}$. A sensitivity improvement of at least one order of magnitude is needed to access more constraining limits on the potential IC emission from WR 25.

\section{Infra-red emission from massive stars}
There is a group of carbon-rich WR stars (called WC stars) that generate thick, dusty circumstellar shells in their winds as seen in their infra-red light curves. Many WCs often undergo variable dust production, some periodic and others random \citep{1995IAUS..163..335W}. Among all the phenomena caused by the collision of stellar winds in early-type binary systems, perhaps the most unexpected is the formation of circumstellar dust. The processes of dust formation by these objects are still not understood, nor are the parameters that determine which of them make dust and which do not. However, the episodic formation of dust by some WR stars indicates that the values of these critical parameters in a particular object can vary so as to start and stop the condensation of dust grains \citep{1991MNRAS.252...49U}. Consequently, the analysis of these variations can provide insight into the  operation of dust-formation processes in WC winds in general, as well as in particular systems. 

An infra-red and X-ray monitoring of WR 125, an episodic dust maker has been carried out. WR 125 (MR 93) is a Galactic WR binary system classified as WC7ed+O9 III with a period of $>$6600 days \citep{2001NewAR..45..135V}. It undergoes mass loss at 6 $\times$ 10$^{-5}$ $M_{\odot}$ yr$^{-1}$ with a terminal velocity of 2900 km s$^{-1}$ (Williams  {\em et al.} 1992). The IR excess of WR 125 started in 1990, lasted for $\gtrsim$ 3 years, being maximum during 1992-93, and also absorption lines were seen in its spectrum supporting its CWB status \citep{1992MNRAS.258..461W,1994MNRAS.266..247W}. No recurrence of the 1990-93 dust formation episode was noticed till 2014 \citep{2014MNRAS.445.1253W}. However, few hints of infra-red brightening of WR 125 at the beginning of 2018 have been provided by \citet{2019MNRAS.488.1282W}.

WR 125 has been explored using Soft X-ray Telescope (SXT) onboard ASTROSAT \citep{2014SPIE.9144E..1SS} in addition to the other X-ray observations obtained with \textit{Einstein}, \textit{ROSAT}, \textit{Swift} and \textit{XMM$-$Newton} during the years 1980--2020. Near-infrared (NIR) observations have been taken with TIRCAM2 mounted on the 3.6-m Devasthal Optical Telescope (DOT; \citealt{2018BSRSL..87...29K,2018JAI.....750003B}). The X-ray emission (0.3-10.0 keV energy range), especially the soft X-rays, is observed to switch to a low state in the year 2020 pointing toward the next periastron passage in WR 125. The drop in the soft X-ray emission could be attributed to a significant photoelectric absorption close to the periastron, when the X-ray emission from the WIR may be more quantitatively absorbed by the dense WC wind. Considering the previous low X-ray emission observed in the year 1991 by \textit{ROSAT}, an orbital period of 28-29 years is suggested for WR 125. 

\begin{figure}[h]
\centering
\includegraphics[scale=0.45]{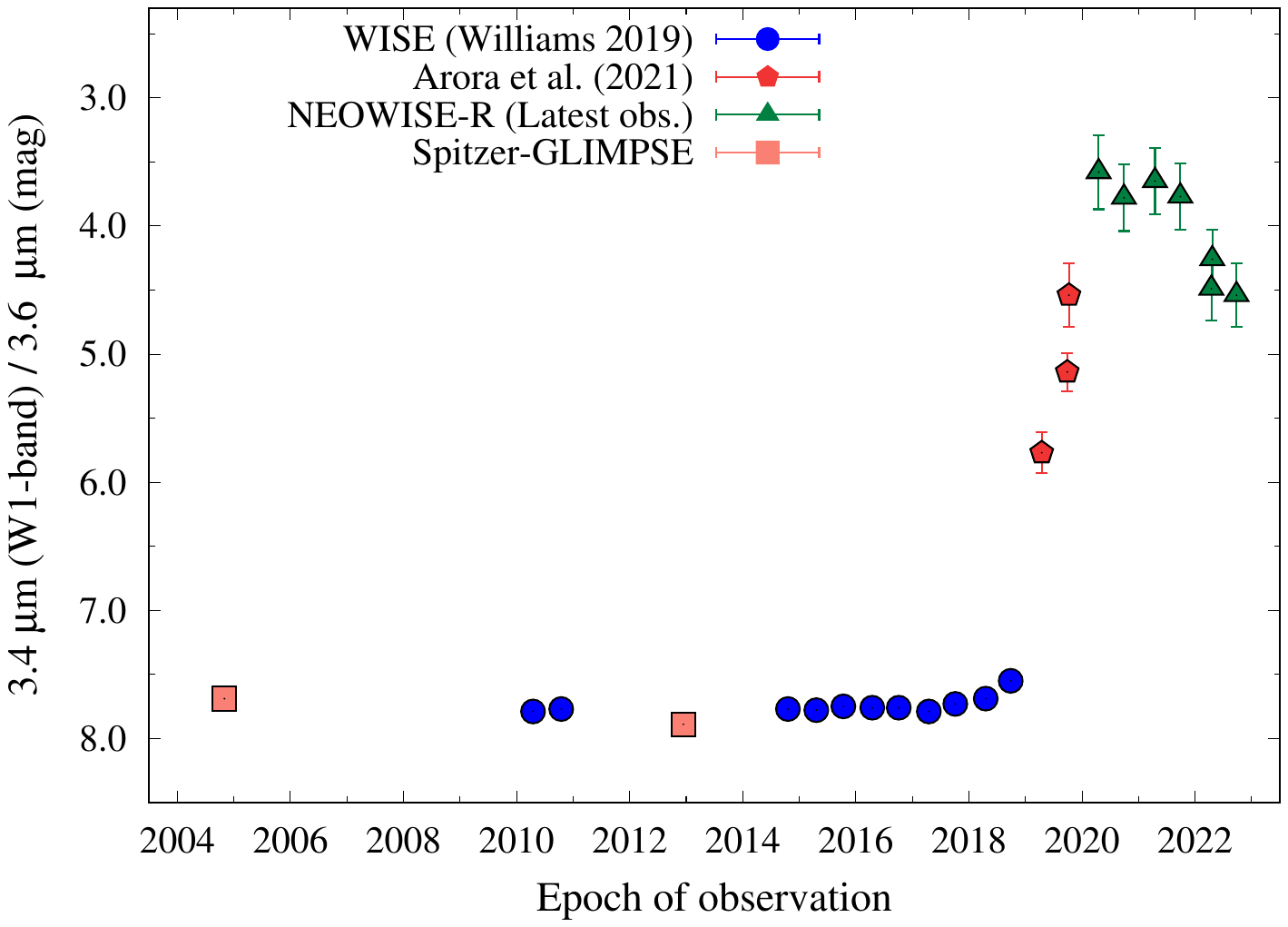}\label{fig4}
\begin{minipage}{12cm}
\caption{MIR light curve of WR 125 created by combining WISE 3.4 $\mu$m (W1-band) data and Spitzer GLIMPSE magnitudes in the 3.6 $\mu$m wave band from the literature and the present study.}
\end{minipage}
\end{figure}

The anticipations drawn from X-ray analysis are further assured by the NIR observations of WR 125 obtained with TIRCAM2. The \textit{K}--band photometric measurements during 2019-2021 showed brightened IR emission by $\sim$0.6 magnitude. Further, the mid-infrared (MIR) light curve of WR 125 was generated using the Near-Earth Object WISE Reactivation (NEOWISE-R; \citealt{2011ApJ...731...53M})  survey  observations in the W1-band (3.4 $\mu$m) and Spitzer GLIMPSE \citep{Churchwell_2009} survey data in the 3.6$\mu$m band.  The principal data set used for this study is the 2023 data release of the NEOWISE-R survey comprising of sky observations taken between 2013--2022. The wavelength of the NEOWISE W1-band is well suited for observing $\sim$1000 K circumstellar dust emission. The continuous survey observations proves most useful for studying variations over few years. The NEOWISE-R data points have been corrected for the saturation effects, as mentioned in \citet{2014ApJ...792...30M}. The latest W1-band observations taken from April, 2020 to September 2022 have been compared with the previously recorded photometric magnitudes during the survey as mentioned in \citet{2019MNRAS.487.2624A} and \citet{2019MNRAS.488.1282W} and it clearly display enhancement around the year 2019--2021 in Figure 4.

The excess infra-red emission is attributed to the circumstellar dust formation close to the periastron passage. It appears that the system started coming out of the periastron passage in the beginning of the year 2022 as the MIR emission just starts to fade out around that time. The wind collision is not sufficiently strong in that part of the binary orbit to power dust formation further so as to replenish the evaporated dust formed earlier. The present infra-red outburst is identical to the one observed in the beginning of the year 1990 that lasted for about 3--4 years around the periastron passage of an eccentric long-period binary. Again, the time interval between the two dust formation episodes is about 28-29 years which corresponds to the orbital period of the binary (see \citealt{2021AJ....162..257A} for details). 

\section{Implications of multiwavelength exploration of massive binaries}
A very important aspect of wind collision in massive star systems is the acceleration of particles up to relativistic velocities by the strong magneto-hydrodynamic shocks. The acceleration process transfers some mechanical energy from the shocks to charged particles (electrons, protons, or even heavier nuclei). When relativistic electrons are present, they can participate in the production of radio synchrotron radiation in the presence of the local magnetic field (mainly due to the massive stars). Such systems are also termed as particle accelerating colliding wind binaries (PACWBs; \citealt{2013A&A...558A..28D}, \url{https://www.astro.uliege.be/~debecker/pacwb/}). Thus, the presence of synchrotron radiation is a tracer of particle acceleration, that only happens in the presence of the shocks of colliding winds in a binary system (for full discussion, see \citealt{DA2023}, this proceeding). 

The binarity/multiplicity investigation of massive stars is necessary to understand their formation as well as the evolution. The census of short-period systems is easier to achieve using classical techniques (e.g. spectroscopy, high-resolution imaging), but for long-period systems (above several months up to decades) strong biases significantly affect the statistics. Therefore, one has to rely on the indirect techniques of detecting long-period binaries. The multiwavelength observations of early-type stars are important to unveil the presence of companions and their orbital parameters through the exploration of strong hydrodynamic shocks, dust formation and particle acceleration in massive star winds. This is based on the combination of measurements in the infrared and the radio domain (ground measurements) and in X-rays (space observations) as has been done for WR 25 and WR 125. In addition to the multiplicity, a multiwavelength approach is necessary to study the dynamics and properties of stellar winds which not only influence the star itself but also affect the surrounding medium at the scale of star-forming regions or galaxies.

\begin{acknowledgments}
This work is supported by the Belgo-Indian Network for Astronomy and astrophysics (BINA), approved by the International Division, Department of Science and Technology (DST, Govt. of India; DST/INT/BELG/P-09/2017) and the Belgian Federal Science Policy Office (BELSPO, Govt. of Belgium; BL/33/IN12).
\end{acknowledgments}



\begin{furtherinformation}

\begin{orcids}
\orcid{0000-0002-1360-4853}{Bharti}{Arora}
\orcid{0000-0002-1303-6534}{M. De}{Becker}
\orcid{0000-0002-4331-1867}{Jeewan C}{Pandey}


\end{orcids}

\begin{authorcontributions}
All authors contributed significantly to the work presented in this paper.
\end{authorcontributions}

\begin{conflictsofinterest}
The authors declare no conflict of interest.
\end{conflictsofinterest}

\end{furtherinformation}

\bibliographystyle{bullsrsl-en}

\bibliography{extra}

\end{document}